\documentclass{article} 
\usepackage{graphicx} 
\usepackage{amsmath}
\usepackage{amsfonts}
\usepackage{amssymb}
\date{\today}
\usepackage{graphicx} 
\usepackage{float}

\newcommand{\insertplot}[5]{\begin{figure}
 \hfill\hbox to 0.05in{\vbox to #5in{\vfill
 \inputplot{#1}{#4}{#5}}\hfill}
 \hfill\vspace{-.1in}
 \caption{#2}\label{#3}
 \end{figure}}
 \newcommand{\inputplot}[3]{
 \special{ps: plotfile #1}
}
\newcounter{fig}

\newcommand{\ee}{\end{equation}}
\newcommand{\eea}{\end{eqnarray}}
\newcommand{\be}{\begin{equation}}
\newcommand{\bea}{\begin{eqnarray}}

\usepackage{xcolor}

\begin{document}

 \title{Boson stars and  black holes with \\
 (complex and) real scalar hair
}

\author{
{\large Yves Brihaye}$^{1}$, {\large Fabien Buisseret}$^{2,3}$,
{\large Betti Hartmann}$^{4}$, and
{\large Oliver Layfield}$^{4}$
\\
\\
$^{1}${\small Physique de l'Univers, Universit\'e de
Mons, 7000 Mons, Belgium}
\\
$^2${\small Service de Physique Nucl\'{e}aire et Subnucl\'{e}aire, Universit\'{e} de Mons, 20 Place du Parc, 7000 Mons, Belgium}\\
$^3${\small CeREF, Chaussée de Binche 159, 7000 Mons, Belgium} \\
$^{4}${\small Department of Mathematics, University College London, Gower Street, London, WC1E 6BT, UK}
}
\maketitle
\begin{abstract}
We discuss boson stars and black holes with scalar hair in a model where the complex scalar field forming the boson star and the hair on the black hole, respectively, interacts with a real scalar field via a H\'enon-Heiles-type potential. We demonstrate that black holes and boson stars carrying only a real scalar field with cubic self-interaction are possible and that black holes with both real and complex scalar field branch off from these solutions for sufficiently large interaction between the two fields and/or sufficiently large horizon radius $r_h$. The latter possess lower mass for the same choice of coupling constants than the former, however seem to be thermodynamically preferred only for high enough temperature.
\end{abstract}

\section{Introduction}

Boson stars are theorised compact objects, the result of gravitationally bound compact collections of bosonic particles which mediate an outward scalar field \cite{Kaup:1968zz,Mielke:1997re,Friedberg:1986tq,Jetzer:1991jr,Schunck:2003kk}.
The  non-gravitating counterparts of these solutions are often referred to as $Q$-balls \cite{Coleman:1985ki}, non-topological solitons formed of a complex scalar field with harmonic time-dependence. These solitons exist only for (non-renormalizable) self-interaction potentials of (at least) sextic order. 
This changes when considering boson stars, where only a mass term for the scalar field is required. Following studies of topological solitons
and the possibility of inserting black holes into the centres of these objects thus constructing {\it hairy} black holes (see e.g. the construction of these
solutions in a model with magnetic monopole solutions \cite{Lee:1991vy,Breitenlohner:1991aa}) it was considered that placing a black hole inside the centre of a boson star might lead to a black hole with complex scalar hair. However, this does not works when the radial pressure associated to the scalar field is larger than the pressure in angular direction and when the weak energy condition is fulfilled \cite{Pena:1997cy}. For the spherically symmetric, non-rotating boson star with energy-momentum content fulfilling the weak energy condition the radial pressure is always larger than the pressure in angular direction and hence black holes cannot carry complex scalar hair. However, spherically symmetric, non-rotating black holes 
with complex scalar hair can e.g. be constructed \footnote{We do not discuss extended gravity models here.} when 
the model possesses a U(1) gauge field \cite{Hong:2020miv,Herdeiro:2020xmb} allowing for the radial pressure to be smaller than the pressure in angular direction. Another possibility is to consider rotating black holes
\cite{Herdeiro:2014goa}. In both of the cases discussed, a so-called {\it synchronisation condition} has to be fulfilled which assigns either the electric potential at the horizon or the horizon velocity to the frequency of the complex scalar field, respectively.

In this paper, we discuss another possibility: black holes which carry interacting complex and real scalar hair that interact via a H\'enon-Heiles potential. In our model, the weak energy condition can be violated for certain choices of the coupling constants. Black holes with real scalar
hair have recently been constructed in a model with a quartic, asymmetric potential \cite{Chew:2022enh} and in a model where a U(1) gauged complex scalar field
interacts with a real scalar field via a potential that is quadratic in the complex and quartic in the real scalar field \cite{Kunz:2023qfg}, respectively.

Here, we will choose a scalar potential which describes the interaction between a real and complex scalar field  inspired by the H\'enon-Heiles potential from classical mechanics \cite{HH}. This potential has at most cubic terms in the fields, depends on two masses and two coupling constants. It has been shown to exhibit solitons \cite{Nugaev:2014ima} and $Q$-ball solutions to this model have been examined as well \cite{BB_HH}. It is through coupling to gravity that we extend these solutions in the present work. 

Our paper is organised as follows: in Section 2, we give the model and the equations of motion resulting from a spherically symmetric Ansatz. In Section 3, we discuss scalarised boson stars, i.e. boson stars that carry a real scalar field as well as solitonic objects made off only a real scalar field, while in Section 4, we discuss the hairy black hole solutions.
We conclude in Section 5.

\section{The model}
We consider the following action that describes a complex scalar field non-minimally coupled to a real scalar field in curved space-time:
\be
\label{eq:action}
 S = \int {\rm d}^{4} x  \bigg[ \frac{R}{16\pi G_N} + {\cal L}_{\rm m}\bigg]
 \ee
 with the matter Lagrangian density given by that of two scalar fields (one complex, one real) interacting~:
 \be
 {\cal L}_{\rm m}=-
 \partial_{\mu}\phi^{*} \partial^{\mu}\phi -
 \frac{1}{2}\partial_{\mu}\xi \partial^{\mu}\xi - U(\vert\phi\vert,\xi) \ \ , \ \  U(\vert\phi\vert,\xi)= m_1^2 \phi^{*} \phi + \frac{1}{2} m_2^2 \xi^2 - g_1 \xi \phi^{*} \phi - g_2 \xi^3 
 \label{eq:lagrangian}
 \ee
 where $R$ is the Ricci scalar, $G$ Newton's constant, $\phi$ the complex-valued scalar field and $\xi$ the real-valued scalar field. $m_1$ and $m_2$ are the masses of the complex and real scalar field, respectively, and $g_1$ and $g_2$ are the interaction and self-interaction couplings, respectively. The original H\'enon-Heiles model is such that $g_2=-g_1/3$ \cite{HH}, but generalized forms have been extensively studied in classical mechanics, see e.g. \cite{conte} for a review of soliton-like solutions in generalized H\'enon-Heiles potentials. Here we assume that all parameters belong to $\mathbb{R}^+_0$.  The model has recently been studied in flat space-time \cite{BB_HH}. The equations that result from the variation of the action (\ref{eq:action}) with respect to the metric and matter fields are the Einstein equation ($\alpha = 4\pi G_N, c=1$  in the following): 
 \begin{equation}
G_{\mu\nu}=\frac{8\pi G_N}{c^4} T_{\mu\nu} = 2 \alpha T_{\mu\nu}
\end{equation}
with energy-momentum tensor $T_{\mu\nu} = g_{\mu\nu} {\cal L}_{\rm m} - \frac{\partial {\cal L}_{\rm m}}{\partial g^{\mu\nu}}$ given by
\begin{eqnarray}
T_{\mu\nu} &=& -g_{\mu\nu}\left[\frac{1}{2}g^{\sigma\rho} \left(\partial_{\sigma} \phi^* \partial_{\rho}\phi + \partial_{\rho} \phi^* \partial_{\sigma}\phi + \partial_{\sigma} \xi \partial_{\rho} \xi\right) + m^2 \phi^* \phi + \frac{1}{2} M^2 \xi^2 -g_1 \xi\phi^* \phi - g_2 \xi^3\right] \nonumber \\
&+& \partial_{\mu} \phi^* \partial_{\nu}\phi + 
\partial_{\nu} \phi^* \partial_{\mu}\phi +
\partial_{\mu} \xi \partial_{\nu}\xi 
\end{eqnarray}
as well as the Klein-Gordon equations for the two scalar fields
\begin{equation}
\left(\Box - \frac{\partial U}{\partial \vert \phi\vert^2}\right)\phi=0 \ \ \ \  , \ \
\ \ \Box \xi - \frac{\partial U}{\partial \xi}=0 \ .
\end{equation}

We are interested in spherically symmetric solutions and hence use spherical coordinates ($t,r,\theta,\varphi)$. The Ansatz then reads
\be
    \phi = e^{i \omega t} \frac{F(r)}{\sqrt{2}} \ \ \ , \ \ \ \xi = G(r),
		\label{eq:ansatz1}
\ee
for the matter fields with $F(r)$ and $G(r)$ real-valued functions, and
\be
{\rm d}s^2 = -N(r) \sigma(r)^2 {\rm d}t^2 + \frac{1}{N(r)} {\rm d}r^2 +r^2 {\rm d}\theta^2 + r^2\sin^2\theta {\rm d}\varphi^2 \ \ \ , \ \ \ N=1-\frac{2m(r)}{r}
\label{eq:ansatz2}
\ee
for the metric. $m(r)$ is the mass functions. The equations that result from the variation of the action with respect to the matter and metric field functions read:

\begin{eqnarray}
N' &=& -\alpha r \left(m_1^2 F^2 + m_2^2 G^2 - g_1 GF^2 - 2 g_2 F^3 + \frac{\omega^2 F^2}{N\sigma^2} + N F'^2 + NG'^2\right) + \frac{1}{r} - \frac{N}{r} \ , \label{eq:N}\\ 
\frac{\sigma'}{\sigma} &=& \alpha r \left( F'^2 + G'^2  + \frac{\alpha r F^2 \omega^2}{N^2 \sigma^2} \right) \ , \label{eq:sigma} \\ 
F'' &=& \frac{1}{N} \left[ \left( \left( m_1^2 - \frac{\omega^2}{N \sigma^2} \right) - g_1 G \right) F - \left( N' + \frac{\sigma' N}{\sigma} + \frac{2N}{r} \right) F' \right] \ , \label{eq:F}\\ 
G'' &=& \frac{1}{N} \left[ \left( m_2^2 - 3 g_2 G \right) G - \frac{g_1 F^2}{2} - \left( N' + \frac{\sigma' N}{\sigma} + \frac{2N}{r} \right) G' \right] \ , \label{eq:G}
\end{eqnarray}
where the prime here and in the following denotes the derivative with respect to $r$. In order to obtain asymptotically flat, finite energy solutions, we need to require the following conditions at infinity:
\be\label{eq:bound_infinity}
 F(r\rightarrow\infty) \rightarrow 0 \ \ , \ \ G(r\rightarrow\infty) \rightarrow 0 \ \ , \ \  \sigma(r\rightarrow \infty)\rightarrow 1  \ .
\ee
\\
Other boundary conditions are determined by the object in question. We examine two cases: globally regular solutions as well as black holes. In the former case boundary conditions will be imposed at $r=0$ and in the black hole case they will be imposed at the horizon $r=r_h$. We have solved the differential equations (\ref{eq:N}) - (\ref{eq:G}) numerically using a collocation method with adaptive grid scheme \cite{colsys, colsys2}. 

\subsection{Physical quantities}
The model is invariant under  a global $U(1)$ symmetry $\phi\rightarrow e^{i\chi}   \phi$, $\chi \in\mathbb{R}$. The corresponding locally conserved Noether current $j^{\mu}$ reads~: 
\begin{equation}
j^{\nu}=i\left(\phi^* \partial^{\nu}\phi - \phi \partial^{\nu}\phi^*\right) = \frac{ \omega F^2 }{N \sigma^2}  \ \  .
 \end{equation}
Hence, the solutions possess a conserved charge $Q_N$ given by 
\begin{equation}\label{charge}
Q _N=  \int  \sqrt{-g} \ j^t {\rm d}^3 x 
= 4 \omega \pi \int_{r_0}^{\infty} \frac {F^2 r^2}{N \sigma} {\rm d}r
 \end{equation}
 where $r_0=0$ for boson stars and $r_0=r_h$ for black holes. This has frequently been interpreted as the number of scalar bosons making up the boson star and the scalar cloud surrounding the black hole, respectively. 

In the probe limit, i.e. when $\alpha=0$, we compute the mass of the solution via the spatial integral of the energy density ${\cal E}=-T^t_t$ as follows:
\begin{equation}\label{energy}
M = -  \int  \sqrt{-g} \ T^t_t {\rm d}^3 x = 4 \pi \int  r^2 \sigma  \left[ \frac{\omega^2 F^2}{2N\sigma^2} + \frac{N}{2} \left(F'^2 + G'^2 \right) + \frac{m_1^2F^2}{2} + \frac{m_2^2 G^2}{2} - \frac{g_1 G F^2}{2} - g_2 G^3 \right] dr
 \end{equation}
while for $\alpha\neq 0$, we use the ADM mass such that $M=M_{\rm ADM}=m_{\infty}/\alpha$ with $m_{\infty}=m(r\rightarrow \infty)$.

For black holes, we can further define thermodynamical quantities. The temperature of a static black hole is  given by $T_{\rm H}=\kappa_s/(2\pi)$ with surface gravity $\kappa_s$ ~:
\be
\kappa_s^2=-\frac{1}{4} \left(g^{tt} g^{ij} \frac{\partial g_{tt}}{\partial x^i}\frac{\partial g_{tt}}{\partial x^j}\right)_{r=r_h} = \left(\frac{N'\vert_{r=r_h} \sigma (r_h)}{2}\right)^2  \ . 
\ee
The entropy $S$ and free energy ${\cal F}$ are given as follows:
\be
S=\frac{A_h}{4} = \pi r_h^2 \ \ , \ \ {\cal F}=M - T_{\rm H}S
\ee
where $A_h$ is the surface area of the horizon.

\section{Scalarized boson stars}
In the following, we will discuss the boson stars that are made up out of the complex scalar field $\phi$. We will show that in our model, these boson stars can carry additional real scalar fields. Next to the boundary conditions (\ref{eq:bound_infinity}) we need to impose  boundary conditions at $r=0$ to ensure that we find globally regular solutions to the equations (\ref{eq:N})-(\ref{eq:G}). These read:
\be
\label{eq:bcBS}
N(0)=1 \ \ ,  \ \ F'(0) = 0\ \ , \ \ G'(0) = 0  \ .
\ee
We set $m_1 = 1, m_2 =2$ and $g_2 = 1$ in the rest of this section. This can
be done without loss of generality because appropriate rescalings of the coordinates and functions can be applied. \\
\\

\begin{figure}[h!]
    \includegraphics[scale=0.8]{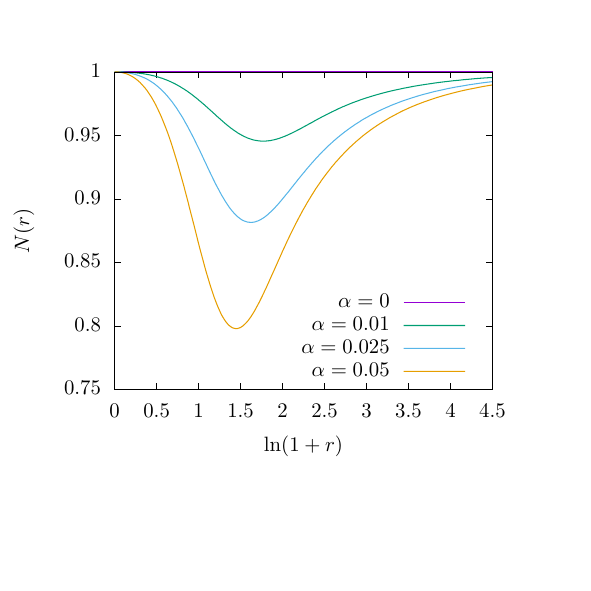}
    \includegraphics[scale=0.8]{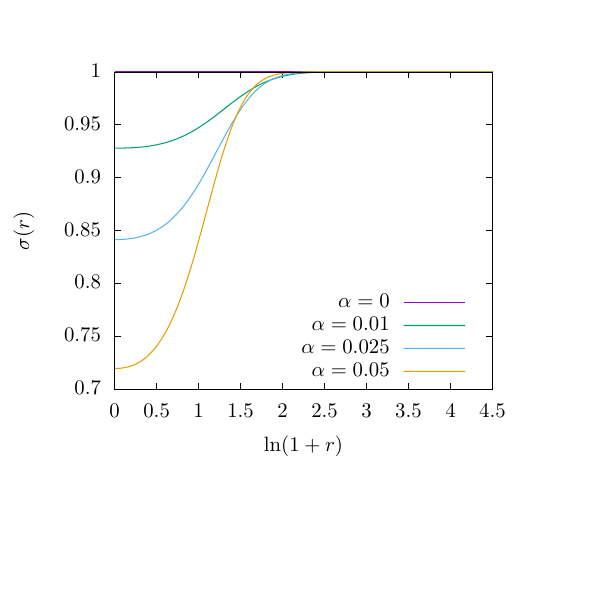}
    
    \vspace{-1.5cm}
    \includegraphics[scale=0.8]{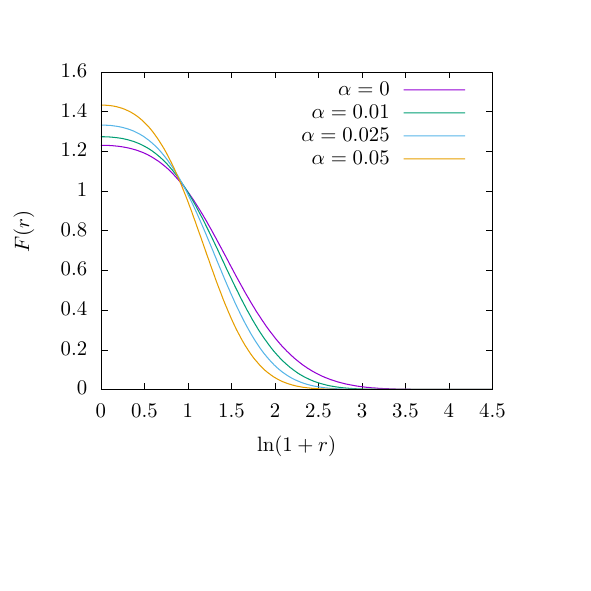}
    \includegraphics[scale=0.8]{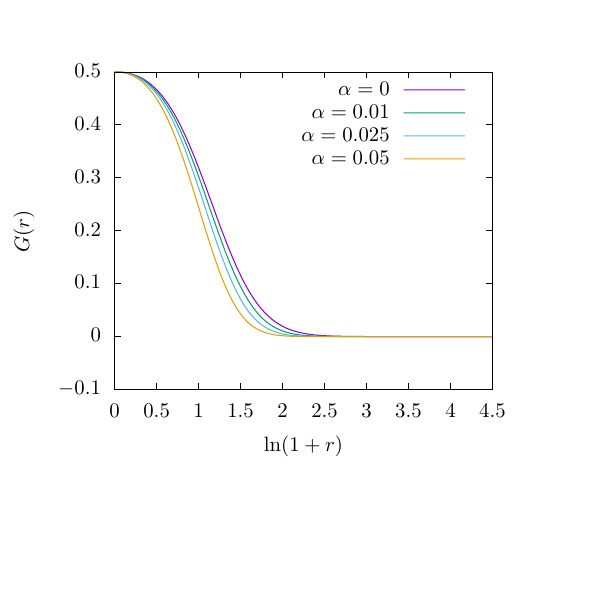}
    
    \vspace{-1cm}
    \caption{Metric functions $N(r)$ (top left), metric function $\sigma(r)$ (top right) as well as the scalar field functions $F(r)$ (bottom left) and $G(r)$ (bottom right), respectively, for $G(0)=0.5$, $g_1=g_2=1$ and different values of the gravitational coupling $\alpha$. }
    \label{fig:typical_solution}
\end{figure}

For $\alpha=0$, the $Q$-ball solutions of this model have been discussed
in detail \cite{BB_HH}. Here, we will focus on investigating the effect
of backreaction, i.e. we will choose different values of $\alpha$ and determine
how the properties of the solutions change. 
In Fig. \ref{fig:typical_solution} we show typical solutions for different
values of $\alpha$. We observe that the minimal value of $N(r)$ as well as the value of the metric function $\sigma(r)$ at $r=0$ decreases when increasing $\alpha$ from zero. We also observe that the scalar field functions become more compact in the sense that the fall-off of the functions happens at smaller $r$. The value of $F(0)$ increases with increasing $\alpha$, while the value of $G(0)$ is fixed in our calculations. The solutions correspond to different values of $\omega$: $\omega= 0.9888$, $\omega=0.9527$, $\omega=0.9107$, and $\omega=0.8525$ for $\alpha=0$, $\alpha=0.01$, $\alpha=0.025$ and $\alpha=0.05$, respectively. This means that keeping $G(0)$ fixed leads to a decrease in the frequency of the complex scalar field when increasing the gravitational backreaction.

 We will now examine the qualitative changes in the boson star solutions  for the three cases of $g_1 = g_2$, $g_1>g_2$ and $g_1<g_2$. The relative value of the couplings $g_1$ and $g_2$ have a clear impact on the system, an observation that has already been made for the corresponding flat space-time solutions \cite{BB_HH}.

\subsection{The case $g_1$ = $g_2$}

\begin{figure}[h!]
    \centering
    \includegraphics[scale=0.8]{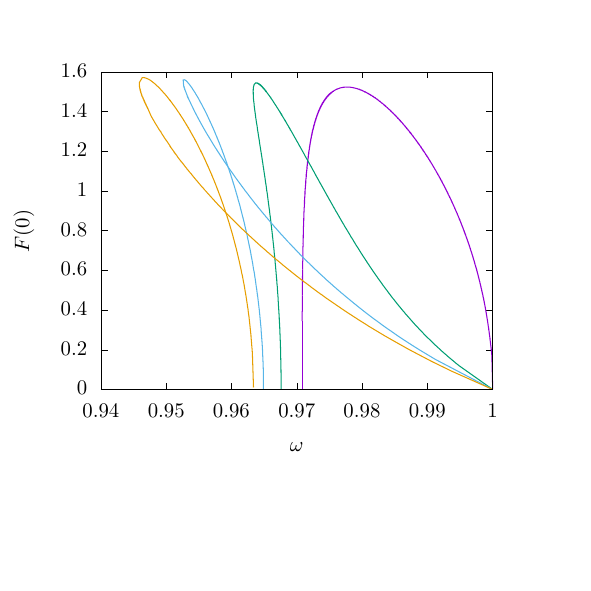}
    \includegraphics[scale=0.8]{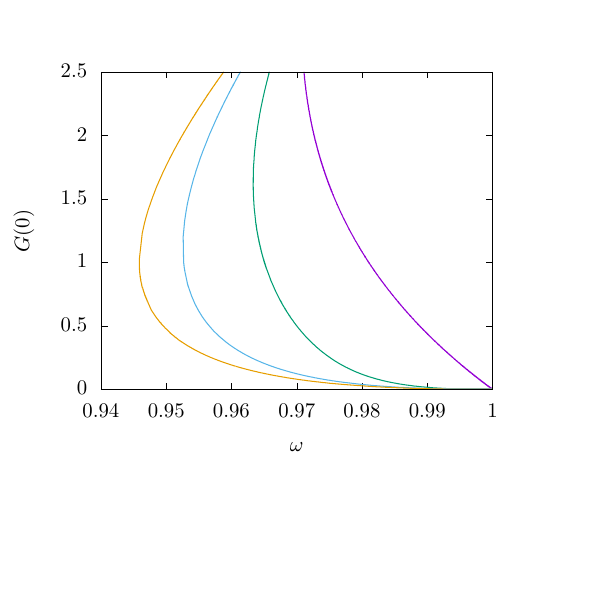}

    \vspace{-1.5cm}
    \includegraphics[scale=0.8]{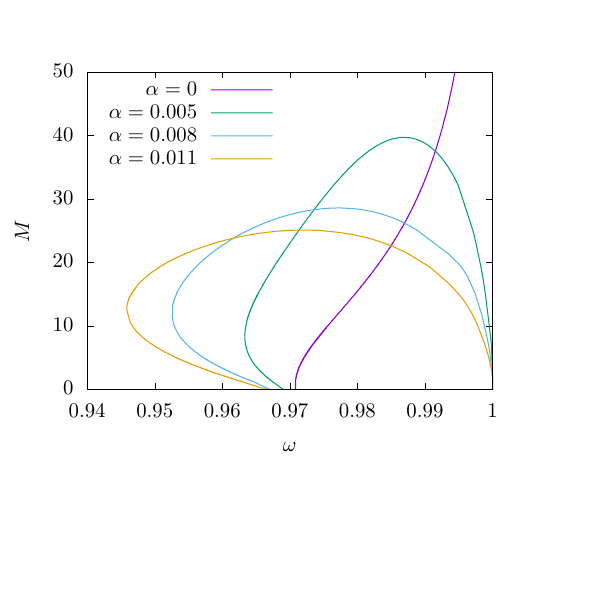}
    \includegraphics[scale=0.8]{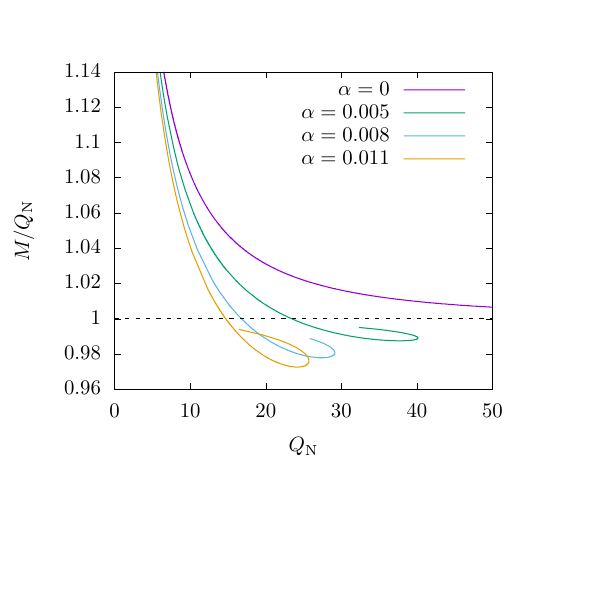}

    \vspace{-1cm}
    \caption{The central value of the complex scalar field $F(0)$ (top left) and the central value of the real scalar field $G(0)$ (top right) in function of $\omega$ for different choices of $\alpha$ and $g_1=g_2=1$ (colour coding as in bottom figures). The mass $M$ in function of $\omega$ (bottom left) and the ratio of the mass and the Noether charge $M/Q_{\rm N}$ (bottom right) in function of the Noether charge $Q_{\rm N}$ for different choices of $\alpha$ and $g_1=g_2=1$.}
    \label{fig:FG_omega}
\end{figure}

\begin{figure}[h!]
    \centering
    \includegraphics[scale=0.8]{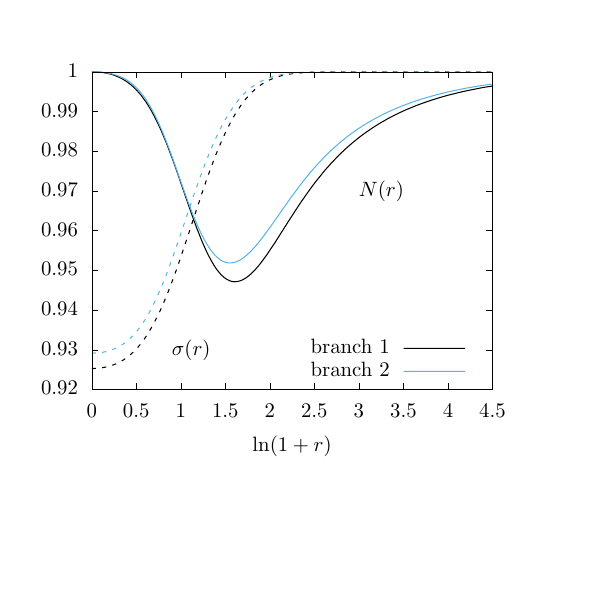}
    \includegraphics[scale=0.8]{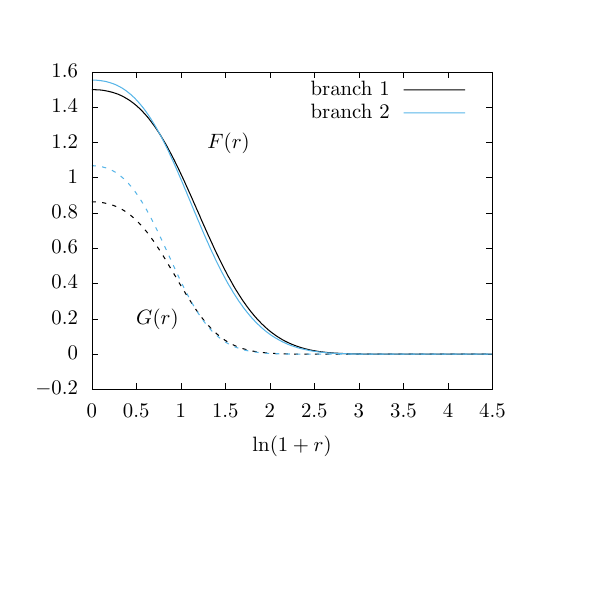}

    \vspace{-1.5cm}
    \caption{We show the two solutions that exist for the same values of the parameters
    with $\alpha=0.011$, $\omega=0.946$ and $g_1=g_2=1$: branch 1 (black) and branch 2 (blue).
    The metric functions $N(r)$ (solid)
    and $\sigma(r)$ (dashed) are shown on the left, while the scalar field functions
    $F(r)$ (solid) and $G(r)$ (dashed) are shown on the right.}
    \label{fig:comparison_solutions}
\end{figure}

We first examine the range of solutions in the $g_1$ = $g_2$ case. Our results are shown in Fig. \ref{fig:FG_omega}, where we give the values of $F(0)$ and $G(0)$ in function of $\omega$. Note that due to the boundary conditions (\ref{eq:bcBS}) imposed $F(0)=0$ implies $F(r)\equiv 0$. For $\alpha=0$, non-trivial solutions
exist on the interval $\omega \in [\omega_{\rm min}, \omega_{\rm max}]$ where the maximal possible value of $\omega=\omega_{\rm max}=1$. We find that for each $\omega$ exactly one solution exists and that the central value of $F(0)$ is maximal at some intermediate value of $\omega$, while $G(0)$ is a strictly increasing function when decreasing $\omega$ with $G(0)=0$ at $\omega_{\rm max}=1$. When increasing $\alpha$ from zero, we observe that while $\omega_{\rm max}=1$, the minimal value of $\omega$, $\omega_{\rm min}$, decreases with increasing $\alpha$. Moreover, the maximal value of $F(0)$ now corresponds to the minimal value of $\omega=\omega_{\rm min}$ and the second branch of solutions reaches $F(0)=0$ at $\omega_{\rm cr} > \omega_{\rm min}$. At the same time, $G(0)$ increases from $0$ at $\omega=\omega_{\rm max}=1$ to $\omega=\omega_{\rm min}$ and then continues to increase on the second branch of solutions when increasing $\omega$ from $\omega_{\rm min}$ to $\omega_{\rm cr}$. $G(0)$ reaches its maximal value when $F(0)=0$ on the second branch of solutions. For increasing $\alpha$ the difference between $\omega_{\rm min}$ and $\omega_{\rm cr}$ increases such that the two branches in $F(0)$ intersect at some value of $\omega$ which is smaller than $\omega_{\rm cr}$ and larger than $\omega_{\rm min}$. We see no such intersection for $G(0)$. In Fig. \ref{fig:comparison_solutions} we show the two solutions that exist for the
same choice of all parameters of the model: $\alpha=0.011$, $\omega=0.946$ and $g_1=g_2=1$. Branch $1$ here refers to the branch that starts at $\omega=1$, while branch $2$ is the branch that ends at $\omega_{\rm cr}$. Clearly, the solutions are different.
The central values of the scalar fields, $F(0)$ and $G(0)$, are smaller for the solutions on branch $1$ as compared to those on branch $2$. Moreover, the minimal value of $N(r)$ as well as the central value of the metric function $\sigma(r)$, $\sigma(0)$, is smaller on branch $1$ as compared to branch $2$. This suggests a stronger curvature of space-time for the solutions
on branch $1$. 

In Fig. \ref{fig:FG_omega} (bottom left), we show the mass $M$ of the solutions in function of $\omega$ for different values of $\alpha$. In the $\alpha=0$ limit, the mass diverges at $\omega=1$, while it tends to zero at $\omega_{\rm min}=\omega_{\rm cr}$. For $\alpha\neq 0$, this changes and the mass
tends to zero at both $\omega_{\rm max}$ as well as at $\omega_{\rm cr}$. This is a well known phenomenon for boson stars that does not change in the presence of the additional real scalar field. In Fig. \ref{fig:FG_omega} (bottom right) we show the ratio $M/Q_N$, which can be thought of as the energy per bosonic particle, in function of $Q_{\rm N}$. We find that for $\alpha=0$, this ratio is always larger than unity and that $Q_{\rm N}$ tends asymptotically to unity from above. When $\alpha > 0$, $Q_{\rm N}$ has a finite maximal value which decreases with increasing $\alpha$. 
We approach the limit $M/Q_{\rm N}=1$ from $M/Q_{\rm N}<1$  with a doubling back of the curve. This is a crucial result. It demonstrates that in our model we have bound states. This indicates that these boson stars are stable with respect to the decay into $Q_{\rm N}$ individual scalar bosons.

\subsection{The case $g_1 \neq  g_2$}
We will first discuss the case $g_1 > g_2 = 1$. In \cite{BB_HH} it was shown that the solutions exist on an interval in $\omega\in [\omega_{\rm min}:\omega_{\rm max}]$ with $\omega_{\rm max}=1$. Increasing $g_1$ from unity it was found that the minimal value of $\omega=\omega_{\rm min}$ decreases, i.e. that a stronger coupling between the complex and real scalar field allows the solutions to exist for smaller values of the frequency $\omega$. We confirm this result
and find that the qualitative pattern does not change significantly when including backreaction.

\begin{figure}[h!]
    \centering
    \includegraphics[scale=0.8]{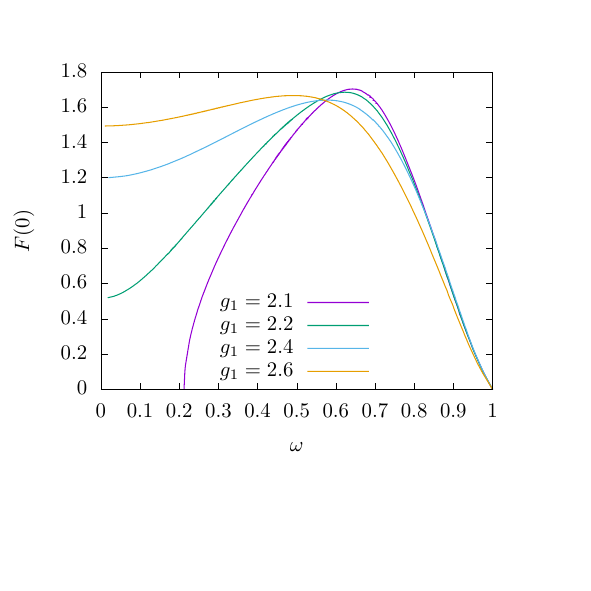}
    \includegraphics[scale=0.8]{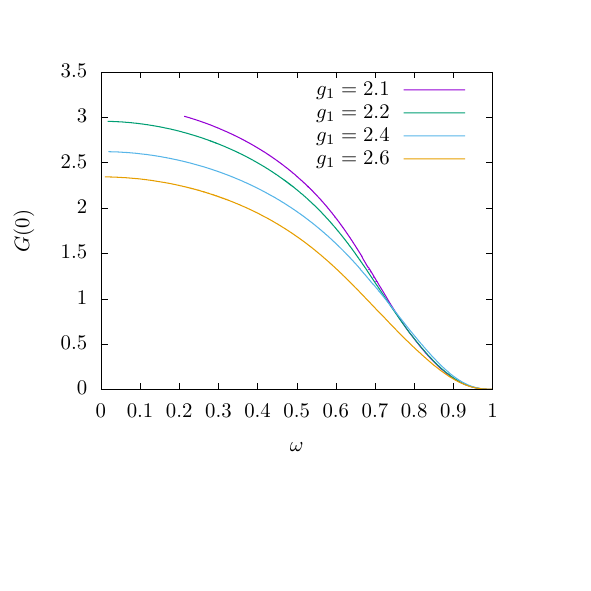}

        \vspace{-1.5cm}
    \caption{{\it Left}: The central values $F(0)$ as function of $\omega$ for $\alpha=0.2$ and several values of $g_1 > g_2=1$. {\it Right}: the corresponding values of $G(0)$. }
\label{fig:g1_larger_g2_alpha02}
\end{figure}

This can be seen in Fig. \ref{fig:g1_larger_g2_alpha02}, where we give $F(0)$ and $G(0)$, respectively, as function of $\omega$ for $\alpha=0.2$ and different values of $g_1$. We find that $\omega_{\rm min}=0$ for sufficiently large $g_1$. The figure demonstrates that while for $g_1=2.1$, $F(0)$ becomes zero at a finite value of $\omega$, the qualitative pattern changes for larger $g_1$. In this case $F(0)=0$ only at $\omega=1$, while
$F(0) > 0$ at $\omega=0$. These limiting solutions have $Q_{\rm N}=0$. Increasing $g_1$ further leads to an increase of $F(0)$ and a decrease of $G(0)$. 
We also find that, as in the $g_1=g_2$ case,  bound state solutions exist.

\begin{figure}[h!]
    \centering
    \includegraphics[scale=0.8]{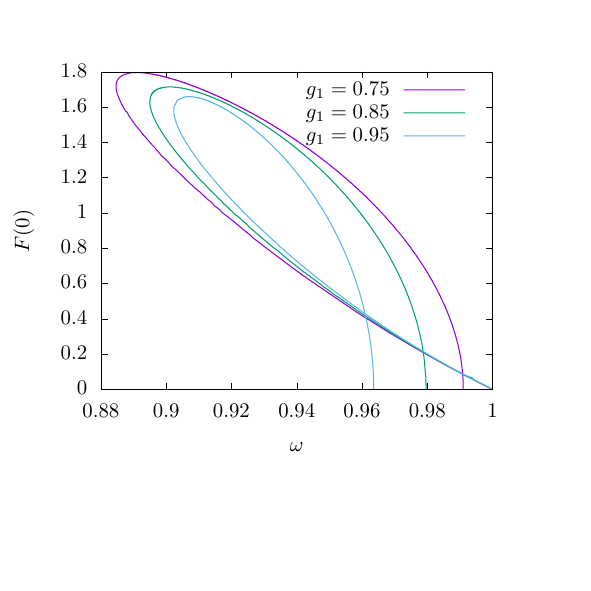}
    \includegraphics[scale=0.8]{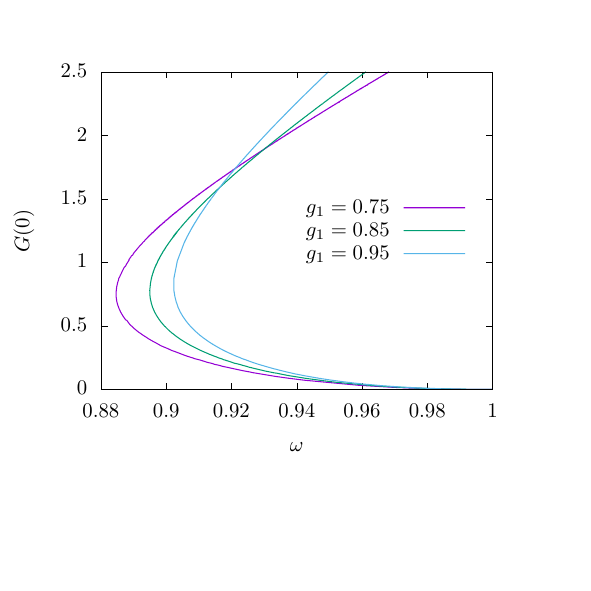}

        \vspace{-1.5cm}
    \caption{{\it Left}: The central values $F(0)$ as function of $\omega$ for $\alpha=0.025$ and several values of $g_1 < g_2=1$. {\it Right}: the corresponding values of $G(0)$. }
    \label{fig:g1smallerg2alpha0025}
\end{figure}

Our results for the case $g_1 < g_2=1$ are shown in Fig. \ref{fig:g1smallerg2alpha0025} for $\alpha=0.025$. Qualitatively similar to the $\alpha=0$ limit, two branches exist in $\omega$ which meet at $\omega_{\rm min}$. We find that $\omega_{\rm min}$ increases with decreasing $g_1$, i.e. solutions exist on smaller intervals of $\omega$ when decreasing $g_1$ from unity. The second branch of solutions always ends at $F(0)=0$
with $G(0)\neq 0$ which is different to the case $g_1 > g_2$. However, we see now also
a crucial difference to the $\alpha=0$ case. We find that increasing $\alpha$ increases both $\omega_{\rm min}$ as well as the maximal value of $F(0)$ for any $g_1 < g_2$. Moreover, decreasing $g_1$ at fixed $\alpha$
 leads to an increase in $\omega_{\rm min}$, which is the opposite of what was observed for $\alpha=0$,  where decreasing $g_1$ decreases $\omega_{\rm min}$.
In comparison to the $g_1 > g_2$ case, the range of $\omega$ for which solutions exist appears more sensitive to $\alpha$ and we were able to obtain solutions for larger values of $\alpha$.

We also find that for $g_1 < g_2$ we can produce bound states and - since we can increase $\alpha$ to larger values - that these bound states are stronger bound as compared to the $g_1  \geq g_2$ case. For $\alpha=0.025$ we find e.g. that the minimal value of $M/Q_{\rm N}$ is approximately $0.92$.

\clearpage

\section{Scalarized Black Holes}
We impose a horizon at $r=r_h>0$ with $N(r_h)=0$. We require all matter fields to be regular at $r_h$ and hence need to impose the following boundary conditions:

\begin{equation}
N'F'\big|_{r=r_h} =\left. \left(m_1^2 - g_1G\right)F\right|_{r=r_h}
\end{equation}

\begin{equation}
N'G'\big|_{r=r_h} = \left.\left(\left(m_2^2 - 3g_2G\right)G - \frac{g_1 F^2}{2}\right)\right|_{r=r_h}
\end{equation}

Moreover, as the scalar field equations demonstrate, we need to choose $\omega=0$ making the complex scalar field real. These three boundary conditions above, along with the three already provided in (\ref{eq:bound_infinity}), give the required six boundary conditions to solve the equations of motion (\ref{eq:N}) $-$  (\ref{eq:G}). There are now six parameters: $m_1$, $m_2$, $\alpha$, $g_1$, $g_2$ and $r_h$. We will comment on appropriate scalings that allow to set some of these values to fixed values without loosing generality in the following.

\subsection{Probe limit}
For $\alpha=0$ the gravity equations have a simple solution:
$\sigma\equiv 1$ and $N(r)=1-r_h/r$. This is the Schwarzschild solution and we will first study the two interacting scalar fields in the background of this space-time.

\subsubsection{The case $F\equiv 0$}
As is easy to see from (\ref{eq:F}), $F(r)\equiv 0$ is a solution to the equations of motion. A first question is therefore whether Schwarzschild black holes can support the real scalar field $G(r)$. The equation for $G(r)$ reads:
\be
   \left(1 - \frac{r_h}{r}\right) \left(G'' +  \frac{2}{r} G'\right) + \frac{r_h}{r^2} G' = m_2^2 G - 3 g_2 G^2 \ \ , 
\ee
which has to be solved subject to the boundary conditions
\be
  G'(r_h) = r_h\left(m_2^2 G(r_h) -  3 g_2 G^2(r_h)\right)              \ \ , \ \ G(r\to \infty) = 0 \ .
\ee
Note that since the coupling constant $g_1$ becomes irrelevant in this case, appropriate rescalings of the function $G(r)$ and of the radial coordinate $r$ allows us to set $g_2=1$ and $m_2=1$ without loosing generality. In the probe limit the only parameter that remains is the horizon radius $r_h$. 

\begin{figure}[h!]
    \centering
    \includegraphics[scale=0.8]{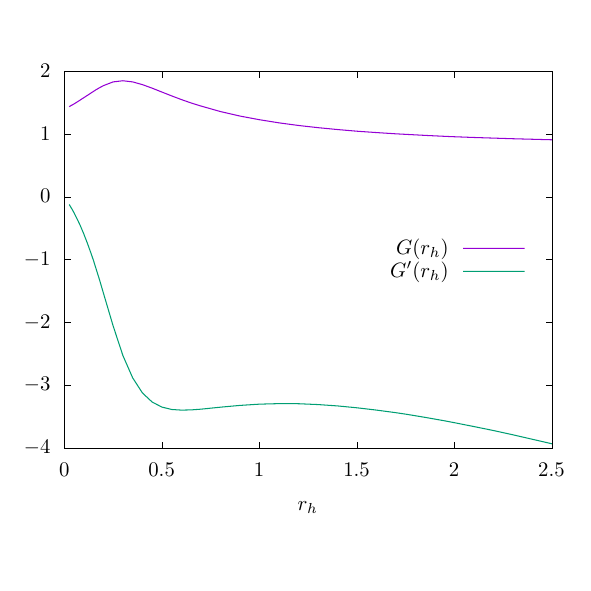}

    \vspace{-0.8cm}
    \caption{We show the values of $G(r_h)$ and $G'(r_h)$ in function of $r_h$ for $\alpha=0$ and $G$-clouds, i.e. solutions with $F(r)\equiv 0$. }
    \label{fig:probe_F_0}
\end{figure}

In spite of its simplicity, we found no closed form solution of this boundary value problem. Analytical results are known for a classical H\'enon-Heiles model in spaces with constant curvature \cite{cHH}, but not in the background of a Schwarzschild black hole. We therefore solved the problem numerically for several values of the horizon value $r_h$. Our results are shown in Fig. \ref{fig:probe_F_0}, where we give the dependence of $G(r_h)$ and $G'(r_h)$ on the horizon radius $r_h$. Since all values are finite, the solutions are regular at $r_h$.
We will refer to these solutions as $G$-clouds in the following. The limit $r_h=0$ is smooth with $G(0)$ finite and corresponds to solutions discussed in \cite{BB_HH}. 

\subsubsection{Two interacting scalar fields}
Let us now consider the case when the two scalar fields are non trivial.

\begin{figure}[h!]
    \centering
    \includegraphics[scale=0.8]{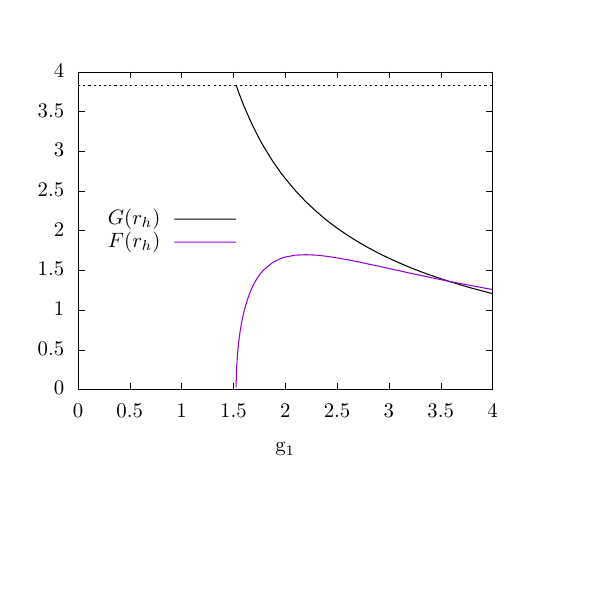}
    \includegraphics[scale=0.8]{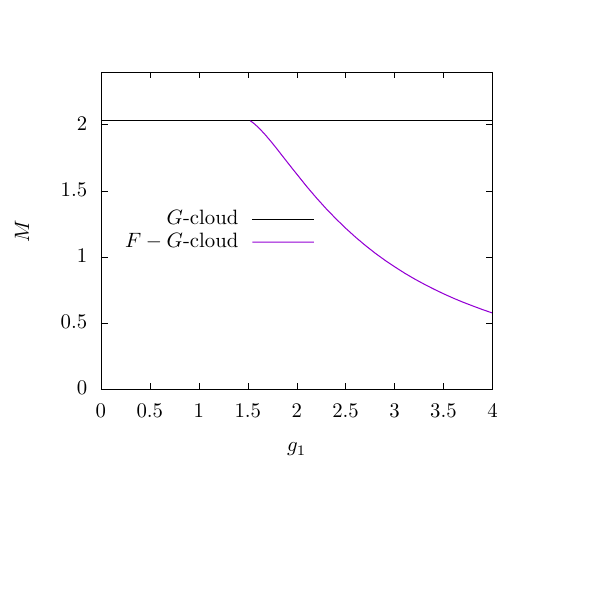}

    \vspace{-1.5cm}
    \caption{{\it Left:} We show the values of $F(r_h)$ (purple) and $G(r_h)$ (black) in function of $g_1$ for $r_h=1$, $m_2=2$, $\alpha=0$. The solid lines correspond to $F$-$G$-clouds, while the dashed lines with constant $G(r)=3.3833$ and $F(r_h)\equiv 0$ correspond to $G$-clouds.
    {\it Right}: We show the values of the mass $M$ in function of $g_1$ for $r_h=1$, $m_2=2$, $\alpha=0$. The constant line at $M\approx 2.0298$ corresponds to the $G$-clouds. }
    \label{fig:probeFGrh1}
\end{figure}

\begin{figure}[h!]
    \centering
    \includegraphics[scale=0.8]{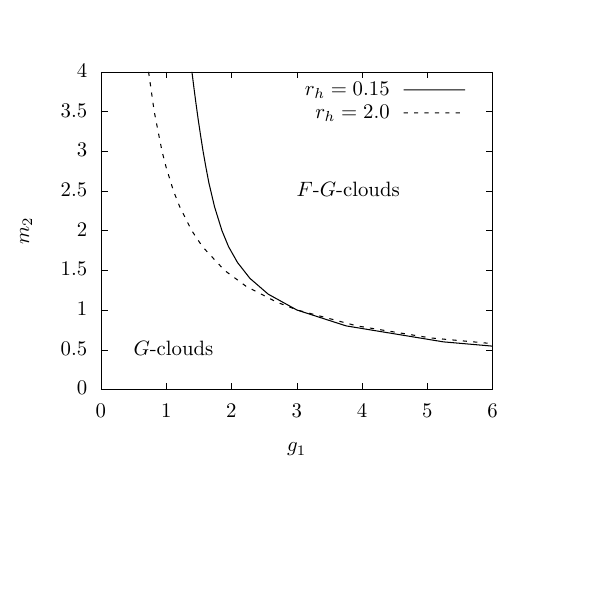}   
    \includegraphics[scale=0.8]{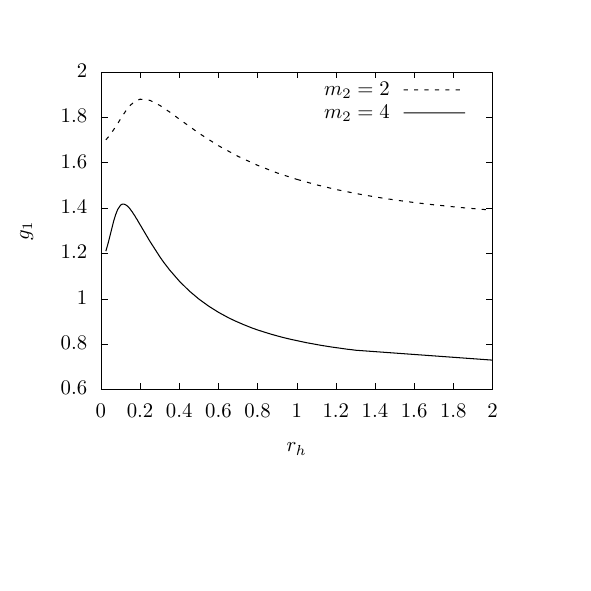}

    \vspace{-1.1cm}
    \caption{{\it Left}: The domain of existence of $G$-clouds and $F$-$G$-clouds in the $m_2$-$g_1$-plane for two different values of $r_h$ for $\alpha=0$, $m_1=2$, $m_2=1$ and $g_2=1$.
    {\it Right}: The domain of existence of $G$-clouds and $F$-$G$-clouds in the $g_1$-$r_h$-plane for two different values of $m_2$, $\alpha=0$, $m_1=2$ and $g_2=1$. The $F$-$G$-clouds exist above the corresponding curves, while $G$-clouds exist in the full domain. In both figures, the curves correspond to the value of $g_{1,cr}$.}
    \label{fig:domain_FGclouds}
\end{figure}

The constant $g_2$ and the mass $m_1$ can be set to unity by appropriate
rescalings of $G(r)$ and of the radial coordinate $r$. The two remaining constants $g_1$ and $m_2$  now play a crucial role for the domain of existence
of the solutions which we will refer to as $F$-$G$-clouds in the following. 
We observe that with fixed $r_h$ and $m_2$ the solutions with $F(r) \neq 0$ exist only for  $g_1 \geq g_{1,cr}$. Our results indeed demonstrate that the function $F(r)$ tends uniformly to the null function for $g_1 \to g_{1,cr}$. In other words: the $F$-$G$-clouds become $G$-clouds in this limit. 
Remembering that the $G$-clouds exist irrespective of the  $g_1$ coupling constant, the $F$-$G$-cloud can be seen
as bifurcating from the $G$-clouds  at the critical value $g_{1,cr}$. Such a bifurcation is shown in Fig. \ref{fig:probeFGrh1} (left) for $r_h=1$, where we give the values of $F(r_h)$ and $G(r_h)$ in function of the horizon radius $r_h$. Note that for $G$-clouds the value of $F(r_h)\equiv 0$  and $G(r_h)\equiv 3.8333$, where the latter results from the choice $m_2=2$. The values $G(r_h)$ for $F$-$G$-cloud solutions are lower than the ones for $G-$cloud. Moreover, the mass $M$ is lower for the $F$-$G$-clouds at fixed horizon radius. This can be seen in Fig. \ref{fig:probeFGrh1} (right) and demonstrates the important role of the {\it a priori} complex scalar field in the system. 
In Fig. \ref{fig:domain_FGclouds} we show the domain of existence of $F$-$G$-clouds in the $m_2$-$g_1$-plane for two different values of $r_h$ (left) and in the $g_1$-$r_h$-plane for two different values of $m_2$ (right). This demonstrates that $g_{1,cr}$ depends on $m_2$ and $r_h$. We find that for small values of $m_2$, the value of $g_{1,cr}$ is not very sensitive to the size (and hence mass) of the black hole. For large(r) values of $m_2$ smaller black holes need larger values of $g_1$ to be able to be surrounded by an $F$-$G$-cloud. We also find that an increase in $m_2$ lowers the value of $g_{1,cr}$ at fixed $r_h$. In summary, we find that the larger the black hole and the larger the mass of the real scalar field, the easier it is to have $F$-$G$-clouds on a Schwarzschild black hole.

\subsection{Black holes with scalar hair}
In the following, we will discuss the case $\alpha > 0$, i.e.
we will study spherically symmetric, static black hole solutions that carry either one or two scalar fields on their horizon and demonstrate how the backreaction of the space-time changes the solutions. We will set $g_2=m_1=1$ without loss of generality in the following. 

\subsubsection{Black holes with $G$-hair}
\begin{figure}[h!]
    \centering
    \includegraphics[scale=0.8]{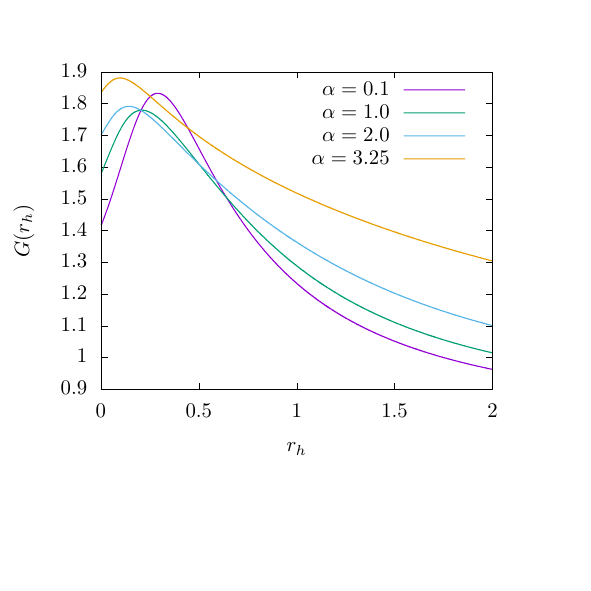}
    \includegraphics[scale=0.8]{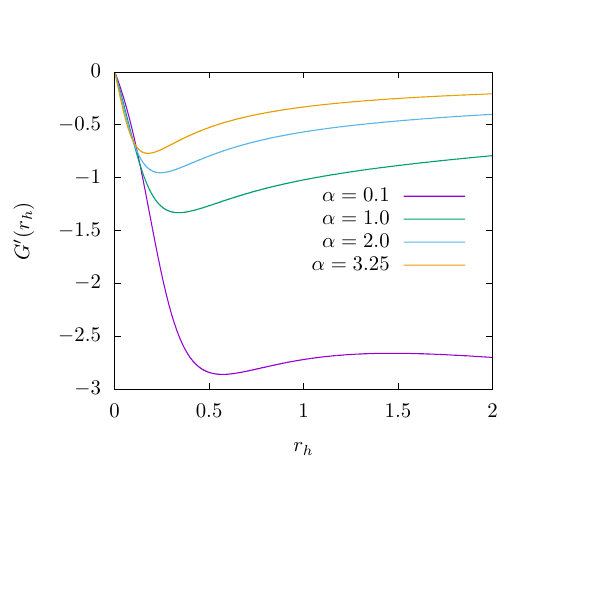}

    \vspace{-1.5cm}
    \caption{We show the values of $G(r_h)$ (left) and $G'(r_h)$ (right), respectively, in dependence on the horizon radius $r_h$ for several values of $\alpha$ and $F(r)\equiv 0$. Moreover $g_1=g_2=1$, $m_1=m_2=1$. }
    \label{fig:Grh_backreacted}
\end{figure}

We will first study the case $F(r)\equiv 0$. We have fixed the value of $\alpha$ and varied $r_h$ for $g_1=m_2=1$.  We show the values of $G(r_h)$ and $G'(r_h)$ in function of $r_h$ in Fig. \ref{fig:Grh_backreacted}. Clearly, $G(r)$ is non-trivial in this case, i.e. we have constructed black holes with $G$-hair. Our numerical data suggests that the solutions are not limited by a maximal value of $r_h$, i.e. we can make the black holes carrying $G$-hair as large as we want. 
The value of $G(r_h)$ increases up to a maximal value at some critical value of $r_h=r_{h,cr}$ from where it monotonically decreases. The corresponding value of $G'(r_h)$ decreases up to
$r_{h,cr}$ and increases from there. We find that increasing $\alpha$ decreases the value of $r_{h,cr}$. Hence, for small black holes, an increase in the size as well as in the backreaction allows for larger values of the real scalar hair on the black hole horizon with scalar field values decreasing for $r > r_h$. For large black holes, the backreaction increases the value of the scalar field on the horizon, while increase in size leads to decrease in the value of the scalar field at $r_h$. Moreover, our results show that for $r > r_{h,cr}$ the scalar
field increases in value when moving away from the horizon. 

\begin{figure}[h!]
    \centering
    \includegraphics[scale=0.8]{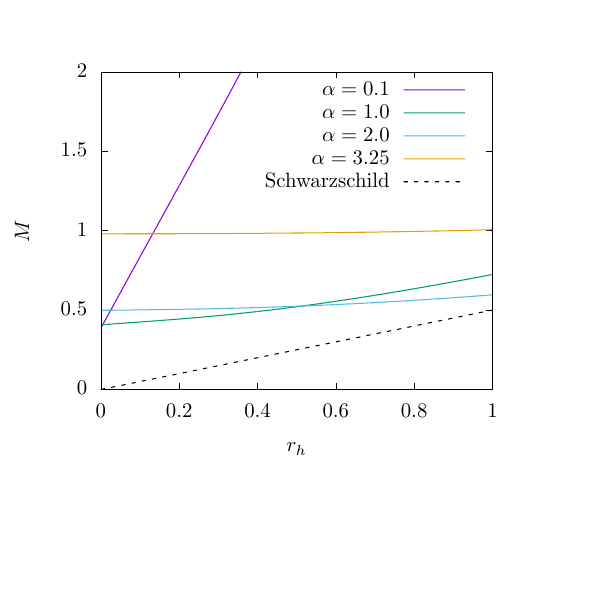}
    \includegraphics[scale=0.8]{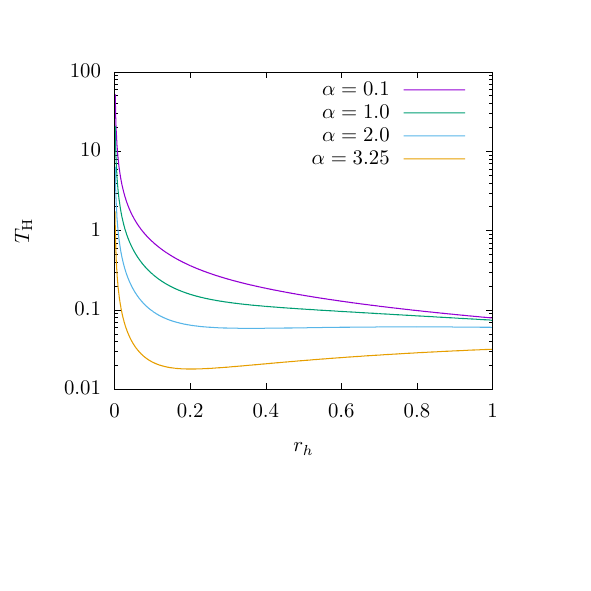}
    
        \vspace{-1.5cm}
    \caption{The values of the ADM mass $M$ (left) and the temperature $T_H$ (right)  in dependence of $r_h$ for several values of $\alpha$ and $F(r)\equiv 0$, $g_1=g_2=1$ and $m_1=m_2=1$.}
    \label{fig:Grh_backreacted_thermo}
\end{figure}

The mass of the solutions increases with increasing $r_h$, see Fig. \ref{fig:Grh_backreacted_thermo} (left), although we find that the slope of the $r_h$-$M$-curve decreases with increasing $\alpha$. For small $\alpha$, we see a strong increase in mass $M$ when increasing $r_h$, while
$M$ seems nearly unchanged for large $\alpha$. At the same time,
the black hole temperature $T_H$ decreases monotonically for increasing $r_h$, see Fig. \ref{fig:Grh_backreacted_thermo}, when $\alpha$ is small. For large $\alpha$, on the other hand,
we find that the temperature seems to possess a local minimum at some intermediate horizon value. 
 In the limit $r_h \to 0$ the temperature diverges 
and we find a regular solution. These are the gravitating versions of the solitons made of real scalar fields first discussed in \cite{BB_HH} for $\alpha=0$. (For soliton solutions in models with more general potentials  see also \cite{PhysRevLett.32.1080}).) We find that for $\alpha=0.1$ we have $G(0)\approx 1.4$, while \cite{BB_HH} finds
$G(0)=1.397$ for $\alpha=0$. Hence, the solitonic solutions made of a real scalar only generalize to curved space-time.

\begin{figure}[h!]
    \centering
    \includegraphics[scale=0.8]{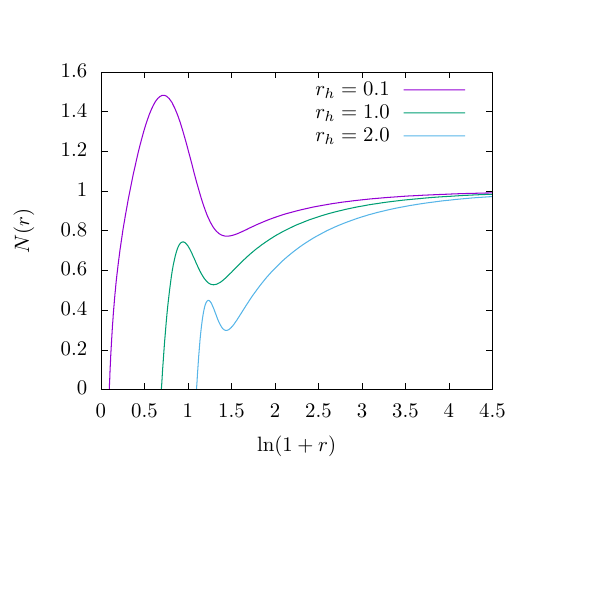}
  \includegraphics[scale=0.8]{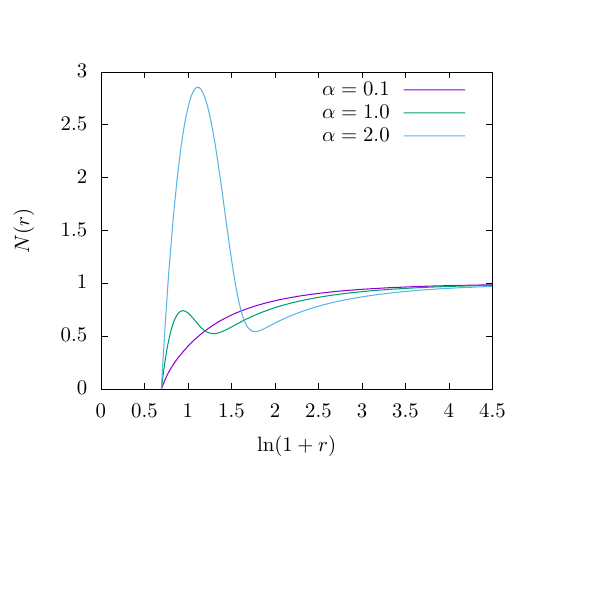}

  \vspace{-1.5cm}
    \caption{{\it Left}: The metric function $N(r)$ of black holes with $G$-hair ($F(r)\equiv 0$) for different values of the horizon radius $r_h$ and $g_1=g_2=1$, $m_1=m_2=1$, $\alpha=1.0$. {\it Right}: 
    The metric function $N(r)$ of black holes with $G$-hair ($F(r)\equiv 0$) for different values of the gravitational coupling $\alpha$ and $g_1=g_2=1$, $m_1=m_2=1$, $r_h=1.0$.
    }
    \label{fig:metricN2}
\end{figure}

Typical profiles of the metric function $N(r)$ are presented
in Fig. \ref{fig:metricN2} (left) for three values of $r_h$, $\alpha=1.0$, $g_1=g_2=1$, $m_1=m_2=1$.  We also show profiles of the metric function for a constant horizon radius $r_h=1$ and several values of $\alpha$ in Fig. \ref{fig:metricN2} (right). We find for all parameter choices presented in these figures that $N(r)$ possesses a local maximum and a local minimum outside the horizon - expect for the choice $r_h=1$, $\alpha=0.1$. The maximal value of $N(r)$ increases, while the minimal value of $N(r)$ decreases when decreasing $r_h$ for fixed $\alpha$. For fixed $r_h$, the minimal value does not show strong dependence on $\alpha$ (although it shifts to large $r$ when increasing $\alpha$), however, the maximal value increases strongly when increasing $\alpha$. We hence find, as expected, the strongest change in the metric function $N(r)$ for small black holes with large backreaction.

\subsubsection{Black holes with $F$-$G$-hair}
Let us finally present the results for black holes carrying both real as well as complex scalar hair, i.e. $F$-$G$- scalar hair. 

\begin{figure}[h!]
    \centering
    \includegraphics[scale=0.8]{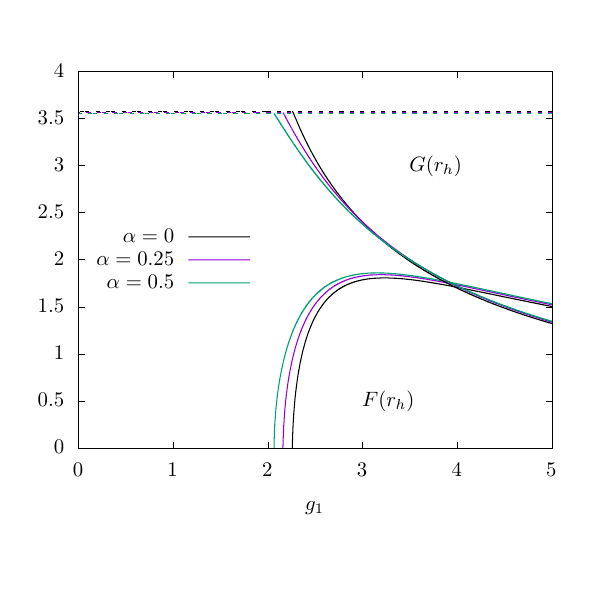}

    \vspace{-1cm}
    \caption{The values of $F(r_h)$ and $G(r_h)$ in dependence of $g_1$ for several values of $\alpha$ with $m_1=m_2=1$, $g_2=1$, and $r_h=0.15$. The solid lines are associated to the black holes with $F$-$G$-clouds, while the dashed lines are those for the black holes with $G$-clouds. The lines at zero or starting from zero at a finite value of $g_1$ represent the values of $F(r_h)$, while the constant lines with non-vanishing, positive values and those starting from these correspond to the values of $G(r_h)$.}
    \label{fig:FGblackholes}
\end{figure}

\begin{figure}[h!]
    \centering
    \includegraphics[scale=0.8]{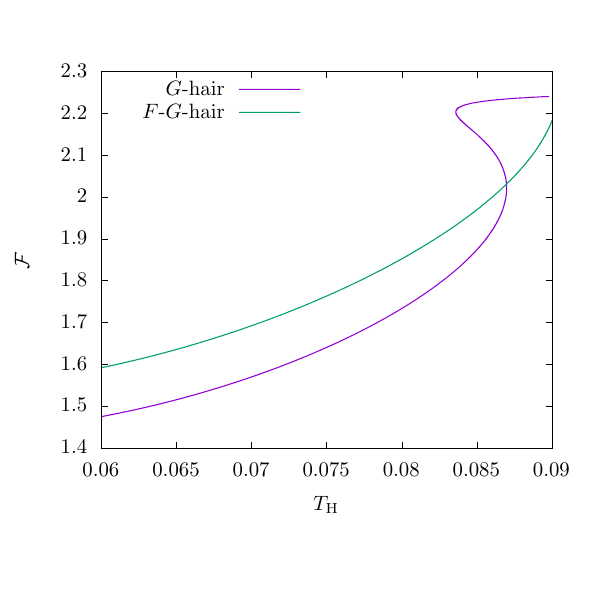}

    \vspace{-1cm}
    \caption{We show the free energy ${\cal F}$ in function of the temperature $T_{\rm H}$ for
    black holes with $G$-hair and with $F$-$G$-hair, respectively. Here $\alpha=0.5$, $m_1=1$, $m_2=2$, $g_1=2.5$, $g_2=1$.}
    \label{fig:free_energy}
\end{figure}

We show the values of $F(r_h)$ and $G(r_h)$ in function of $g_1$ for different values of $\alpha$ with $r_h=0.15$, $m_2=1$ and $g_1=1$ in Fig. \ref{fig:FGblackholes}.  Here we see that increasing the gravitational strength lowers the value of $g_1$ at which solutions with $F$-$G$-hair bifurcate from those with only $G$-hair. In Fig. \ref{fig:free_energy} we show the free energy ${\cal F}$ in function
of the temperature $T_H$ for black holes with $G$-hair and black holes with $F$-$G$-hair for $\alpha=0.5$, $m_2=2$ and $g_1=2.5$.
For low temperature, the black hole with only $G$-hair is thermodynamically
preferred, while at sufficiently high temperature the black holes with $F$-$G$-hair have lower free energy.

\clearpage

\section{Conclusions}

We have extended the work of \cite{BB_HH} to include gravity. We have found globally regular as well as black hole solutions. The globally regular solutions are boson stars (made of the complex scalar field) which carry an additional real scalar field. In comparison to the non-backreacted case we find that now two solutions exist for the same values of the couplings and the frequency of the complex scalar field. Moreover, there exist solutions that are stable with respect to the decay into the individual bosonic particles that make up the star. 
Interestingly, globally regular solutions made solely out of a real scalar field can exist in our model as was already shown in \cite{BB_HH} in flat space-time. While this seems to be in disagreement with Derrick's theorem \cite{derrick} which forbids localized, finite energy solutions made out of static, real scalar fields in three spatial dimensions, one of the requirements, namely that the potential be positive definite, is not fulfilled in our model for appropriate choices of the coupling constants. Hence these solutions can exist and do not contradict the theorem.

Interestingly, we also have constructed static, spherically symmetric and asymptotically flat black holes with scalar hair (real or real and complex). This is not in disagreement with the theorem given in \cite{Pena:1997cy} which states that static, spherically symmetric, asymptotically
flat black hole space-times with energy-momentum content fulfilling the weak energy condition as well as
$T_{\theta}^{\theta} \geq T^r_r$ are necessarily trivial. While the latter condition is always fulfilled in our case, the former is not: ${\cal E}=-T^t_t$ is not necessarily positive definite as the scalar field potential becomes negative for specific choices of
the self-couplings.

To our knowledge, such results have never obtained before. The Hénon-Heiles potential has been shown to emerge as an effective potential for a test particle in a Schwarzschild geometry perturbed by quadrupolar and octupolar terms \cite{Vieira:1996zf,Vieira:1996pv}, but dynamical fields have never been associated to this potential in curved spacetime.\\

%
\bibliographystyle{unsrt}
\bibliography{BSBHHH.bib}

\end{document}